\documentclass[letter, structabstract]{aa}
\usepackage{natbib}
\usepackage{txfonts}
\usepackage{graphicx}

\begin{document}

\title{The Cepheid mass discrepancy and pulsation-driven mass loss}

\author{Hilding R. Neilson \and Matteo Cantiello \and Norbert Langer}
\titlerunning{The Cepheid mass discrepancy and pulsation-driven mass loss}
\authorrunning{Neilson et al.}
\institute{Argelander-institut f\"{u}r Astronomie, Universit\"{a}t Bonn, Auf dem H\"{u}gel 71, 53121 Bonn, Germany}
\date{}
\abstract{A longstanding challenge for understanding classical Cepheids is the Cepheid mass discrepancy, where theoretical mass estimates using stellar evolution and stellar pulsation calculations have been found to differ by approximately $10$ - $20\%$.}
{We study the role of pulsation-driven mass loss during the Cepheid stage of evolution as a possible solution to this mass discrepancy.}
{We computed stellar evolution models with a Cepheid mass-loss prescription and various amounts of convective core overshooting.  The contribution of mass loss towards the mass discrepancy is determined using these models, }
{Pulsation-driven mass loss is found to trap Cepheid evolution on the instability strip, allowing them to lose about $5-10\%$ of their total mass when moderate convective core overshooting, an amount consistent with observations of other stars, is included in the stellar models.}
{We find that the combination of moderate convective core overshooting and pulsation-driven mass loss can solve the Cepheid mass discrepancy. }


\maketitle

\section{Introduction}\label{sec:intro}
The pulsation properties of Cepheids are tightly correlated to their fundamental parameters, such as mass and luminosity, which makes them valuable tools for distance measurements and cosmology. Cepheids are also  powerful probes of stellar evolution thanks to the coupling of stellar evolution and stellar pulsation models, both constraining the internal structure of these stars \citep[e.g.][]{Hofmeister1964, Cox1966, Christy1966}.  However, mass predictions using each method do not agree, \cite{Stobie1969} found that stellar evolution models predict Cepheids have higher masses than do stellar pulsation models for the same effective temperature and luminosity.  This Cepheid mass discrepancy has been a challenge for stellar evolution and pulsation theory for the past $40$ years.

\cite{Cox1980} showed that the mass discrepancy, defined as the mass difference relative to the predicted stellar evolution mass, is approximately $40\%$. \cite{Moskalik1992} claimed that the updated \cite{Iglesias1990} opacities provide a resolution to the mass discrepancy.  However, the current status of the Cepheid mass discrepancy is $17 \pm 5\%$ \citep{Keller2002, Caputo2005, Keller2006, Keller2008}.  Furthermore, there is evidence that the mass discrepancy is a function of both mass \citep{Caputo2005} and metallicity \citep{Keller2006}.

Dynamic masses have been determined for four Galactic Cepheids that are in binary systems: SU Cyg \citep{Evans1990}, V350 Sgr \citep{Evans1997}, S Mus \citep{Evans2006}, and Polaris \citep{Evans2008}.  The dynamic masses are all lower than masses predicted using stellar evolution theory and consistent with stellar pulsation models.  Furthermore, \cite{Pietrzynski2010}  determine the mass of the Large Magellanic Cloud Cepheid OGLE-LMC-CEP0227, which is in an eclipsing binary system, to a precision of $1\%$ and find that it agrees with the mass predicted by stellar pulsation.  \cite{Cassisi2011} model the evolution of this Cepheid and find agreement with the dynamic mass when extra mixing is included. These results suggest that there are physics missing in the stellar evolution calculations.

The two most likely solutions to the mass discrepancy are convective core overshooting in a Cepheid's main-sequence progenitor \citep{Chiosi1992} and mass loss during the Cepheid stage of evolution \citep{Bono2006}.   Convective core overshooting during main sequence evolution mixes extra hydrogen into the core. This leads to a more massive post-main sequence helium core, hence to a more luminous Cepheid  or conversely to a less massive Cepheid for the same luminosity if overshooting is not included in the stellar evolution models.  Overshooting is also required to explain observations of eclipsing binary stars \citep[e.g.][]{Sandberg2010, Clausen2010}, $\beta$ Cephei stars \citep{Lovekin2010} and massive B-type stars \citep{Brott2011}.

On the other hand, mass loss during the Cepheid stage of evolution acts to reduce the stellar mass without affecting the stellar luminosity. \cite{Deasy1988} determined mass-loss rates of $10^{-9}$ to $10^{-8}~M_\odot~\rm{yr}^{-1}$ from IRAS observations.  More recently \citet[][and references therein]{Merand2007} observed infrared excess in nearby Galactic Cepheids from interferometric observations, while Spitzer observations also detected infrared excesses \citep{Marengo2010, Marengo2011, Barmby2010}.  \cite{Neilson2009b, Neilson2010} modeled the infrared excess in Large Magellanic Cloud Cepheids in the OGLE-III \citep{Soszynski2008} and SAGE \citep{Meixner2006} surveys.  In these works, the observed infrared excess was modeled by a dusty wind, suggesting that Cepheids may be undergoing significant mass loss.  

From a theoretical perspective, \cite{Neilson2008, Neilson2009} developed a prescription for pulsation-driven mass loss in Cepheids. They predicted mass-loss rates up to $10^{-7}~M_\odot~\rm{yr}^{-1}$. While this evidence suggests Cepheid mass loss is important, it remained unclear whether enough mass is lost during the Cepheid stage of stellar evolution to account for the measured Cepheid mass discrepancy.

The purpose of this work is to test whether pulsation-driven mass loss in Cepheids is an important contributor towards solving the Cepheid mass discrepancy.  In the next section, we estimate the order-of-magnitude change in mass that may occur during the Cepheid stage of evolution due to mass loss based on the \cite{Neilson2008} prescription.  In Sect.~\ref{sem}, we compute detailed stellar evolution models to explore the role of mass loss and convective core overshooting. In Sect.~\ref{dis}, we summarize our results.

\section{Analytic test of pulsation-driven mass loss}
We estimate the amount of mass loss during the Cepheid stage of evolution, where we assume the Cepheid lifetime is equivalent to the helium-burning timescale $\tau_{\rm{He}}$ for a given stellar mass.

We compute an average Cepheid mass-loss rate by assuming that the Cepheid instability strip is infinitesimally thin. Thus, for a given mass and luminosity there is only one value for the pulsation period, amplitude of luminosity variation, and amplitude of radius variation.  The mass-loss rate is then computed using the pulsation-driven mass-loss prescription developed by \cite{Neilson2008}, who hypothesized that pulsation in the envelope of a Cepheid generates shocks that carry momentum to the surface of the star and enhances the mass loss that an evolved star undergoes.  Mass-loss rates were found to be enhanced by up to three orders of magnitude. In that prescription, the mass-loss rate is a function of the stellar mass, luminosity, radius, pulsation period, and pulsation amplitudes of the luminosity and radius. We determine the pulsation period using the bolometric Leavitt law \citep{Turner2010}, and the radius is given by the period-radius relation \citep{Gieren1989, Neilson2010}.  The amplitudes of the velocity and brightness variation are computed from period-amplitude relations \citep{Klagyivik2009}.  The change in radius is the period times the velocity amplitude, and we assume the $V$-band amplitude is equivalent to the bolometric change of brightness. We note that the predicted pulsation amplitudes have significant errors, but the only other way to predict these amplitudes is using nonlinear pulsation models. Thus, given a stellar mass and luminosity, we determine the period, radius, and pulsation amplitudes, hence the mean mass-loss rates.

The Cepheid's luminosity is determined from stellar evolution calculations.  We use the \cite{Heger2000} stellar evolution code to compute models with masses $M = 4$ - $9~M_\odot$ in steps of $1~M_\odot$.  The models are computed by assuming no convective core overshooting in the main sequence progenitors and no pulsation-driven mass loss during the post main-sequence evolution.  We also determine the helium-burning timescales for each mass from the models.
We use the predicted timescales $\tau_{\rm{He}}$ and pulsation-driven mass-loss rates $\dot{M}$ to determine the contribution of mass loss towards the mass discrepancy, as shown in Table.~\ref{tab1}. The range of mass-loss rates is consistent with the results of \cite{Neilson2008}.  It should be noted that the lower mass-loss rate for the $6~M_\odot$ model is due to a local minimum in the period-amplitude relation at a period $\lesssim 10.4$ day.  These results suggest, however, that mass loss is not a solution for the Cepheid mass discrepancy of $17\pm 5\%$.  The change in mass relative to the initial mass for the low-mass Cepheids appears consistent with the discrepancy but for the high-mass ($8,9~M_\odot$) Cepheids, there is a negligible mass change due to pulsation-driven mass loss.  Furthermore, the helium-burning timescale is longer than the Cepheid lifetime, meaning the mass changes presented are upper limits.
\begin{table}[t]
\caption{Contribution to the mass discrepancy due to pulsation-driven mass loss.}\label{tab1}
\begin{center}
\begin{tabular}{cccc}
\hline
\hline
Mass & $\tau_{\rm{He}}$ & $\dot{M}$ & $\Delta M/M$  \\
 $(M_\odot)$ & (Myr) & $(M_\odot~\rm{yr}^{-1})$ &  (\%)\\
\hline
4 & $31.294$& $1.4\times 10^{-8}$ & $11$\\
5 & $17.487$ &$6.4\times 10^{-8}$& $22$\\
6 & $10.088$& $2.6\times 10^{-9}$ & $4.3$\\
7&  $6.383$ & $1.3\times 10^{-7}$ & $12$\\
8 & $4.867$ &$3.3\times 10^{-8}$ &$2.0$\\
9 &$3.605$ & $4.1\times 10^{-9}$&$0.1$\\
\hline
\end{tabular}
\end{center}
\end{table}

This estimate suggests that mass loss may be a significant contributor but it cannot account for the entire mass discrepancy.  However, detailed calculations are required to explore the feedback of pulsation-driven mass loss on the evolution of Cepheids, which in turn, can affect the contribution of mass loss to the mass discrepancy.

\section{Stellar evolution models with pulsation-driven mass loss}\label{sem}
\begin{figure*}[t]
\begin{center}
\includegraphics[width=\textwidth]{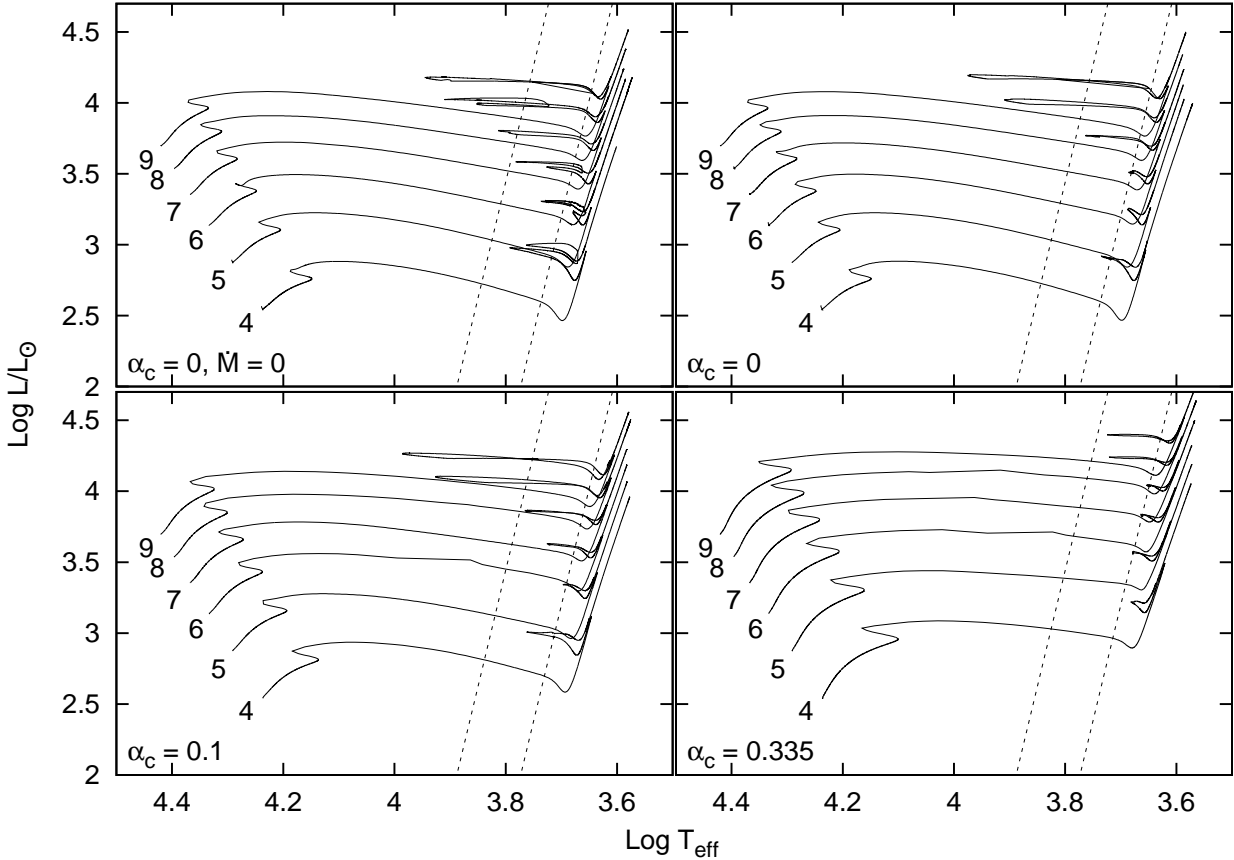}
\end{center}
\caption{Stellar evolution tracks of stars with initial masses from $4$-$9~M_\odot$ (see labels) from the zero age main sequence until core helium exhaustion, computed using the \cite{Heger2000} code for different assumptions of convective core overshooting and mass loss during the Cepheid stage of evolution.  The dashed lines represent the boundaries of the Cepheid instability strip.}\label{fig1}
\end{figure*}

We computed stellar evolution models for masses $M=4,5,6,7,8,$ and $9~M_\odot$ for four different scenarios.  The first scenario is for models with zero convective core overshooting and no pulsation-driven mass loss.  The second, third, and fourth cases include pulsation-driven mass loss and convective core overshooting with $\alpha_c = 0, 0.1$, and $0.335$, respectively.  The $\alpha_c = 0.335$ case is based on the results of \cite{Brott2011}. In the code, convective core overshooting is given by the distance that  convective cells penetrate above the core on an evolutionary timescale, defined as $\Lambda = \alpha_c H_P$,  where $H_P$ is the pressure scale height and $\alpha_c$ a free parameter.   Our stellar evolution tracks are shown in Fig.~\ref{fig1}.  The tracks, especially during blue loop evolution, are very sensitive to the physical processes implemented in the models, including both convective core overshooting and mass loss \citep{Salasnich1999}.

In the models, we assume pulsation-driven mass loss occurs during the Cepheid stage of evolution only, where the blue edge of the Cepheid instability strip is based on the results of \cite{Bono2000}, and we assume that the red edge is parallel to the blue edge with a somewhat arbitrary width.  This is because the location of the red edge on the HR diagram is difficult to define both observationally and theoretically \citep{Fernie1990,Fiorentino2007}.  However, the chosen location of the red edge will not change our main results. Instead of using the Leavitt law to determine the pulsation period, we now employ a period-mass-radius relation from \cite{Gieren1989}, where the uncertainty of the predicted period is about $25\%$. Again, the pulsation amplitudes are computed using the period-amplitude relations of \cite{Klagyivik2009}.  All other fundamental parameters describing a Cepheid are taken from the evolution models.  

The blue loop evolution of each model differs for each scenario.  When pulsation-driven mass loss is included, the width of the blue loops decreases for masses $M < 7~M_\odot$. This phenomenon was found previously by \cite{Brunish1987}, who argued that enhanced mass loss can trap a Cepheid in the instability strip.

\begin{figure}[t]
\begin{center}
\includegraphics[width=0.47\textwidth]{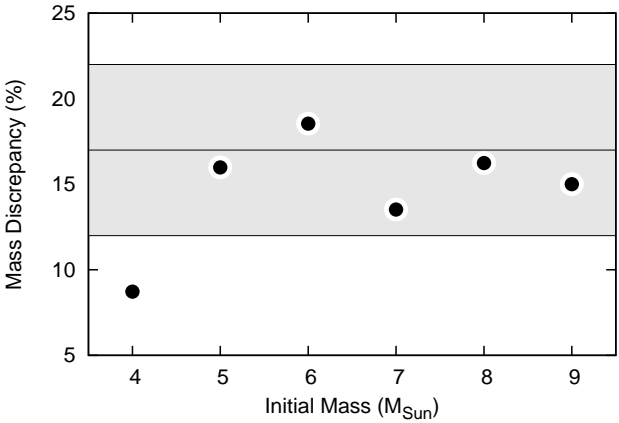}
\end{center}
\caption{Predicted mass difference due to $\alpha_c=0.335$ and pulsation-driven mass loss as a function of initial stellar mass.  The horizontal lines and gray region represent the average mass discrepancy found by \cite{Keller2008}.}\label{fig2}
\end{figure}

We can compute the contribution to the mass discrepancy for the stellar evolution models.  This arises from the contribution to the mass discrepancy due to convective core overshoot plus the relative change in stellar mass due to mass loss, as shown in Table~\ref{t2} for the three cases with pulsation-driven mass loss included. A Cepheid has a mass discrepancy contribution of $2.5\%$ for $\alpha_c = 0.1$ and $8.375\%$ for $\alpha_c = 0.335$ from \cite{Keller2008}. For the case of $\alpha_c = 0.335$ with pulsation-driven mass loss, the predicted contribution to the mass discrepancy, shown in Fig.~\ref{fig2}, is consistent with the results of \cite{Keller2008}.   This suggests that pulsation-driven mass loss in Cepheids is a significant contributor towards the mass discrepancy, while allowing for a smaller amount of convective core overshooting, $\alpha_c = 0.25$ - $0.40$, as opposed to the $\alpha_c = 0.5$ - $1$ suggested by \cite{Keller2008}.  The value for $\alpha_c$ is also consistent with constraints on convective core overshooting in other types of stars \citep{Clausen2010, Lovekin2010, Sandberg2010, Brott2011}.

Most evolution models appear to spend most of the Cepheid lifetime near the assumed red edge of the instability strip. At this location on the Hertzsprung-Russell diagram, the period-amplitude relation overestimates the pulsation amplitude. The relation has an intrinsic dispersion caused by the range of periods and amplitudes a Cepheid may have as it crosses the instability strip for a given stellar luminosity. Thus the amplitude is an averaged value for a given luminosity. On the other hand, nonlinear pulsation models suggest that pulsation amplitudes are largest near the middle of the instability strip \citep{Bono2000}.  \cite{Neilson2008} argued that the pulsation-driven mass-loss rate increases with decreasing pulsation period or increasing pulsation amplitudes, thus suggesting that mass-loss rates at this location are also overestimated. Our simplified approach therefore seems to lead to a situation where the amount of total mass loss and potentially the width of a Cepheid blue loop depend on the assumed location of the red edge of the instability strip. However, we argue that this is not the case.  To show this, we computed a  $6~M_\odot$ stellar evolution model with $\alpha_c = 0.335$, assuming that the instability strip is about half as wide as shown in Fig.~\ref{fig1}.  The contribution to the mass discrepancy, in this case, is approximately the same as before.  This is because a star spends most of its Cepheid lifetime at the tip of the blue loop, which is close to the helium burning timescale, and the mass-loss rate at the tip of the blue loop is approximately the same in both cases.

\cite{Cassisi2011} find that stellar evolution models can match the mass of the Large Magellanic Cloud Cepheid OGLE-LMC-CEP0227, as measured as part of an eclipsing binary system, if one includes convective core overshooting with $\alpha_c = 0.2$ using the definition of the mass discrepancy from \cite{Keller2008}.  This suggests that the mass discrepancy for this particular Cepheid is $5\%$, contrary to the conclusions of \cite{Cassisi2011} who assume that evolution models with moderate convective core overshooting are ``standard'' models.  We also note that the result of $\alpha_c = 0.2$ applies for only one low-mass Cepheid and may not apply for the higher mass Cepheids studied by \cite{Keller2008}. This result is similar to the results shown in Table~\ref{t2} for our $4~M_\odot$ models, where the mass discrepancy is found to vary from about $6\%$ to $8\%$. While this is currently the most precise Cepheid mass known, it is still only one Cepheid, and more precise dynamic masses are needed to constrain the Cepheid mass discrepancy. 

\begin{table}[t]
\caption{Predicted Cepheid mass discrepancy due to pulsation-driven mass loss and different amounts of convective core overshooting as functions of initial mass, $M_{\rm{i}}$. }\label{t2}
\begin{center}
\begin{tabular}{cccc}
\hline
\hline
$M_{\rm{i}}$&\multicolumn{3}{c}{mass discrepancy ($\Delta M/M$)} \\ \cline{2-4}
$(M_\odot)$  &$\alpha_c= 0$ &$\alpha_c= 0.1$&$\alpha_c= 0.335$\\
\hline
$4$ &$6.25\%$&$6.00\%~(3.50\%)$ & $ 8.72\%~(0.34\%)$ \\
$5$ &$0.20\%$&$6.10\% ~(3.60\%)  $&$15.98\%~( 7.60\%)$\\
$6$ &$3.00\%$&$8.30\%~(4.80\%)$ & $ 18.54\%~(10.16\%)$ \\
$7$ &$5.57\%$&$9.07\%~(6.57\%) $&$ 13.52\%~(5.14\%)$ \\
$8$ &$1.75\%$&$4.50\%~(2.00\%)$ & $ 16.24\%~(7.86\%)$ \\
$9$ &$1.67\%$&$4.40\%~(1.90\%)$ & $ 15.00\%~(7.62\%)$\\
\hline
\end{tabular}
\tablefoot{Values in parenthesis denote the change in mass due to mass loss. For the $\alpha_c = 0$ case the change of mass loss is the predicted mass discrepancy.}
\end{center}
\end{table}

The results for the contribution towards the mass discrepancy found in this section differ from those in Sect.~2.  The contribution from mass loss tends to increase with increasing values of $\alpha_c$ because the mass-loss rate depends on the ratio of the luminosity to mass which increases with $\alpha_c$.  Furthermore, an increased luminosity means a larger radius, hence a longer predicted pulsation period.  A longer predicted period suggests larger pulsation amplitudes, which again, results in higher mass-loss rates.  The results in Sect.~2 do not include overshooting, so they miss any feedback from overshooting onto the mass loss.

\section{Summary}\label{dis}
We have shown that pulsation-driven mass loss during the Cepheid stage of stellar evolution based on the \cite{Neilson2008} prescription explains a significant portion of the Cepheid mass discrepancy but not the entire measured discrepancy.  The remaining discrepancy can be explained by convective core overshooting, with a value of $\alpha_c$ that is consistent with measurements in eclipsing binary stars, $\beta$ Cephei stars, and early B-type main sequence stars.

The structure and the width of the model blue loops are also found to be affected by the amount of Cepheid mass loss. For models with higher values of $\alpha_{c}$, the mass-loss rates appear to be high enough to affect the width of the blue loop, which in turn increases the contribution of mass loss to the Cepheid mass discrepancy.  It is this apparent trapping that causes the differences between the predicted mass loss contribution towards the mass discrepancy shown in Sects.~2 and 3.

While our results suggest a possible resolution to the mass discrepancy, more observations of dynamic masses of Cepheids are needed to constrain this theory.

\acknowledgements
HRN is grateful for financial support from the Alexander von Humboldt Foundation.  We would like to thank Dr. Nancy Evans for helpful conversations regarding the dynamic masses of Galactic Cepheids.

\bibliographystyle{aa}
\bibliography{cep}

\end{document}